\documentclass{aa}

\usepackage{graphicx}
\usepackage{txfonts}
\usepackage{caption}
\usepackage{subcaption}
\usepackage{amsmath}	% Advanced maths commands
\usepackage{ulem}
\usepackage{multirow}
\usepackage[bookmarks=false]{hyperref}
\usepackage{natbib}
\usepackage{xcolor}

\usepackage{orcidlink}

\newcommand\Msun{\mathrm{M}_\odot}
\newcommand\Zsun{\mathrm{Z}_\odot}
\newcommand\Mstar{\mathcal{M}_\star}
\newcommand\Mhalo{\mathcal{M}_\mathrm{h}}

\defcitealias{grudic22}{Paper~I}
\defcitealias{rodriguez23}{Paper~II}

\begin{document}

   \title{Great Balls of FIRE III: Modeling Black Hole Mergers from Massive Star Clusters in Simulations of Galaxies}

   \subtitle{}

   \author{Tristan Bruel\inst{1}\fnmsep\thanks{E-mail:tristan.bruel@oca.eu}\orcidlink{0000-0002-1789-7876}
          \and
          Carl L. Rodriguez\inst{2}\orcidlink{0000-0003-4175-8881}
          \and
          Astrid Lamberts\inst{1,3}\orcidlink{0000-0001-8740-0127}
          \and
          Michael Y. Grudi\'c\inst{4}\thanks{NASA Hubble Fellow}\orcidlink{0000-0002-1655-5604}
          \and
          Zachary Hafen\inst{5}\orcidlink{0000-0001-7326-1736}
          \and
          Robert Feldmann\inst{6}\orcidlink{0000-0002-1109-1919}
          }

   \institute{Laboratoire Lagrange, Universit\'e C\^ote d'Azur, Observatoire de la C\^ote d'Azur, CNRS, Bd de l’Observatoire, 06300, France
             \and
             Department of Physics and Astronomy, University of North Carolina at Chapel Hill, 120 E. Cameron Ave, Chapel Hill, NC, 27599, USA
             \and
             Laboratoire Artemis, Universit\'e C\^ote d'Azur, Observatoire de la  C\^ote d'Azur, CNRS, Bd de l’Observatoire, 06300, France
             \and
             Carnegie Observatories, 813 Santa Barbara St, Pasadena, CA 91101, USA
             \and
             Department of Physics and Astronomy, University of California Irvine, Irvine, CA 92697, USA
             \and
             Institute for Computational Science, University of Zurich, Winterthurerstrasse 190, 8057 Zurich, Switzerland
             }

   \date{Received ; accepted }

% \abstract{}{}{}{}{} 
% 5 {} token are mandatory
 
  \abstract
   {After the nearly hundred gravitational-wave detections reported by the LIGO-Virgo-KAGRA Collaboration, the question of the cosmological origin of merging binary black holes (BBHs) remains open. The two main formation channels generally considered are from isolated field binaries or via dynamical assembly in dense star clusters. 
   Here, we focus on understanding the dynamical formation of merging BBHs within massive clusters in galaxies of different masses.
   To this end, we apply a new framework to consistently model the formation and evolution of massive star clusters in zoom-in cosmological simulations of galaxies. Each simulation, taken from the FIRE project, provides a realistic star formation environment with a unique star formation history and hosts realistic giant molecular clouds that constitute the birthplace of star clusters. Combined with the code for star cluster evolution \texttt{CMC}, we are able to produce populations of dynamically formed merging BBHs across cosmic time in different environments.
   As the most massive star clusters preferentially form in dense massive clouds of gas, we find that, despite their low metallicities favourable to the creation of black holes, low-mass galaxies contain few massive clusters and therefore have a limited contribution to the global production of dynamically formed merging BBHs.
   Furthermore, we find that massive clusters can host hierarchical BBH mergers with clear identifiable physical properties. 
   Looking at the evolution of the BBH merger rate in different galaxies, we find strong correlations between BBH mergers and the most extreme episodes of star formation. Finally, we discuss the implications for future LIGO-Virgo-KAGRA gravitational wave observations.}


   \maketitle
%
%-------------------------------------------------------------------

\begin{figure*}
    \centering
    \includegraphics{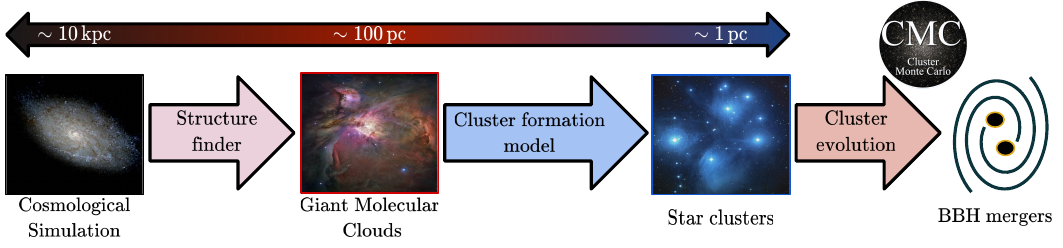}
    \caption{Diagram of our procedure for modelling the formation and evolution of star clusters in a FIRE-2 cosmological zoom-in simulation of a galaxy. The method is described in full in Section \protect{\ref{sec:method}}.}
    \label{fig:cartoon}
\end{figure*}

\section{Introduction}

\begin{table*}
    \centering
    \begin{tabular}{c|c|c|c|c|c}
    %\hline
        \multirow{2}{*}{Simulation} & $\Mhalo$ & $\Mstar$ & \multirow{2}{*}{Number of GMCs} & \multirow{2}{*}{Number of clusters} & \multirow{2}{*}{Number of BBH mergers}\\
         & [$\Msun$] & [$\Msun$] & & & \\
        \hline
        m11i & $7.8\times 10^{10}$ & $9.2\times 10^{8}$ & $3.48\times10^4$ & $3.49\times10^4$ & 1 \\
        m11q & $1.6\times 10^{11}$ & $6.1\times 10^{8}$ & $1.73\times10^4$ & $3.90\times10^4$ & 0 \\
        m11e & $1.7\times 10^{11}$ & $1.4\times 10^{9}$ & $4.13\times10^4$ & $5.87\times10^4$ & 87 \\
        m11h & $2.1\times 10^{11}$ & $3.6\times 10^{9}$ & $1.09\times10^5$ & $1.21\times10^5$ & 280 \\
        m11d & $3.2\times 10^{11}$ & $3.9\times 10^{9}$ & $8.54\times10^4$ & $1.85\times10^5$ & 2208 \\
        m12i & $1.2\times 10^{12}$ & $6.7\times 10^{10}$ & $8.52\times10^5$ & $4.05\times10^6$ & 12438 \\
        %\hline
    \end{tabular}
    \caption{Properties of the primary halo/galaxy at redshift $z=0$ of all the zoom-in simulations used in this study. $\Mhalo$ is the total mass of gas, star, and dark matter contained in the sphere inside which the mean matter density is 200 times the mean matter density of the Universe. The galaxy stellar mass $\Mstar$ is calculated as $90\%$ of the stellar mass within $20\mathrm{kpc}$ of the halo center. The selection criterion of massive clusters $M_\mathrm{cl}\geq6\times10^4\ \Msun$ is only applied at the epoch the cluster is formed.
    }
    \label{tab:sims}
\end{table*}

Since the first direct detection of gravitational waves (GWs) \citep{gw150914} the number of observed compact binary coalescences has been increasing rapidly with each new observing run. The LIGO-Virgo-KAGRA Collaboration (LVK) has reported 90 binary black hole (BBH) merger candidates in the third gravitational wave transient catalog \citep[GWTC-3,][]{gwtc3}, as well as more than 50 additional significant detection candidates to date in the O4 public alerts\footnote{See LVK Public Alerts at \url{https://gracedb.ligo.org/superevents/public/O4/}}. This catalog has provided valuable information about the population of observed BBHs, including their component mass distributions and local merger rate density \citep{gwtc3_pop}. One important question that arises is understanding the various processes that lead to the formation of these BBHs and the environments in which they occur.

Several formation pathways have been suggested \citep[see e.g.][for recent reviews]{mapelli20, mandel22}, with two main categories being favoured: isolated evolution of a stellar binary \citep[e.g.][]{bethe98, belczynski02, dominik12, belczynski16, eldridge16, stevenson17, giacobbo18, neijssel19, santoliquido21, briel22} and dynamical assembly in dense environments such as globular clusters (GCs), young star clusters (YSCs) and nuclear star clusters \citep[e.g.][]{sigurdsson93, zwart00, rodriguez15, rodriguez16, banerjee17, diCarlo20}. Multiple studies have attempted to estimate the contribution of each scenario to the observed population of BBHs \citep[e.g.][]{sedda20, zevin21, wong21, bouffanais21a, mapelli22}, but no consensus has yet been reached on the value of this mixing fraction or on the nature of the dominant channel. However, all these studies agree that multiple formation channels are likely required to explain the properties of the BBH mergers observed by the LIGO and Virgo interferometers, and that isolated and dynamical channels likely contribute at the same order of magnitude.

The usual approach to studying how BBHs are produced in the Universe is to numerically construct cosmological populations of BBHs, whose characteristics can then be compared with the physical properties of the observed BBH mergers. These population synthesis techniques require two fundamental ingredients: a set of initial conditions, and evolutionary models. It is particularly important to use models that are consistent with each other in these two aspects in order to be able to compare populations simulated with different formation channels. A number of codes for star cluster evolution now share the same prescriptions as binary population synthesis suites for stellar evolution and binary interactions \citep[see e.g.][]{diCarlo19, antonini20, rodriguez22}. This ensures similar stellar evolutionary tracks between the two models. In contrast, consistency between the initial conditions of isolated stellar binaries and star clusters seems more challenging to achieve. Yet these initial conditions are of crucial importance. 

Using binary population synthesis methods combined with large sets of star formation models, several studies have shown that the star formation history, and more specifically the metallicity distribution over redshift, has a major impact on the merger rate of BBHs formed in isolated binaries \citep[see e.g.][]{neijssel19, briel22, chruslinska22, broekgaarden22, santoliquido22, srinivasan23}. Regarding the formation and evolution of star clusters in a cosmological context, a number of key questions remain open \citep[see e.g.][for a review]{forbes18}. When and where did they form? What was their initial mass function? While a number of studies have used semi-analytical models to compute a cluster formation rate density as a function of time \citep[see e.g.][]{fall01, rodriguez15, antonini20, mapelli22}, an alternative approach based on the formation of GCs in gas-rich galaxy mergers extracted from large cosmological simulations is also commonly adopted \citep[see e.g.][]{muratov10, li14, choksi18, elBadry18, rodriguez18}. This collection of different initial conditions results in a large uncertainty when estimating the local BBH merger rate \citep{mapelli20}.

Motivated by the idea of using more complex and realistic star formation histories and to investigate certain properties of the host galaxies of double compact objects, various studies have used cosmological simulations as the basis for their population synthesis models of stellar binaries \citep[see e.g.][]{shaughnessy16, lamberts18, lamberts19, mapelli18, artale19, santoliquido22}. These studies all show that the BBHs that merge in the local Universe most probably formed in metal-poor low-mass galaxies, and that they can merge in these small galaxies as well as in larger galaxies after a series of galaxy mergers. Combining a cosmological simulation with a binary population synthesis code can be fairly straightforward, as star particles are directly present in the simulation. In contrast, simulations of galaxies or larger volumes currently do not have a resolution high enough to resolve the formation of star clusters and more complex methods are required \citep[see e.g.][]{lahén19, ma20}. In \citet{grudic22} (hereafter \citetalias{grudic22}), a new framework for modeling cluster formation was applied to a magneto-hydrodynamic simulation of a Milky Way (MW) like galaxy, with a particular focus on the population of young massive clusters. A representative sample of these clusters were then integrated forward in time with the Cluster Monte Carlo code \citep[\texttt{CMC},][]{pattabiraman13, rodriguez22} to create an evolved population of GCs. A detailed description of this study as well as a comparison of the physical properties of these GCs with observations in the MW and M31 are presented in \citet{rodriguez23} (hereafter \citetalias{rodriguez23}).

\begin{figure*}
    \centering
    \includegraphics{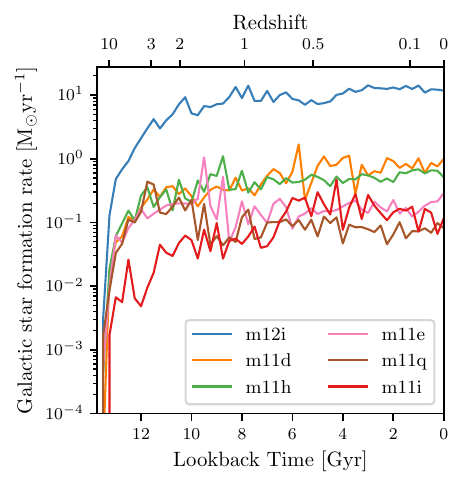}
    \includegraphics{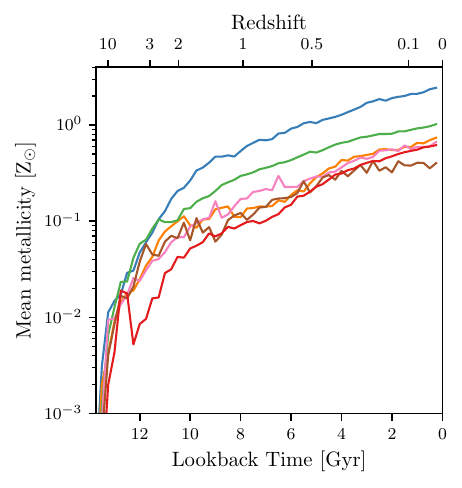}
    \caption{Star formation rate (\textbf{left}) and mean metallicity of newly-formed stars (\textbf{right}) as a function of time in our 6 FIRE-2 cosmological simulations. The star formation rate and mean metallicity are calculated using time bins of 250 Myr.
    }
    \label{fig:sims}
\end{figure*}

In this work, we elaborate upon the \textit{Great Balls of FIRE} series of papers \citepalias{grudic22, rodriguez23}. We extend the analysis to a larger number of galaxies by applying the cluster formation framework to a set of zoom-in cosmological simulations of dwarf galaxies and examine in detail the populations of dynamically formed merging BBHs in the most massive star clusters in each galaxy. Throughout this study, we will only consider BBHs that merge before redshift $z=0$. The term "BBHs" should be interpreted as "merging BBHs" in all that follows.

After describing in detail the cosmological simulations used throughout this study and the methods used to implement the formation and evolution of star clusters in them (Section~\ref{sec:method}), we highlight our results on the populations of massive star clusters in different galaxies and on the BBH mergers that they host (Section~\ref{sec:results}).
We discuss some implications of our results and outline possible extensions of this work (Section~\ref{sec:discussion}), before summarizing our main conclusions about the connection between BBH mergers and their galactic environments (Section~\ref{sec:conclusions}).

\section{Method}\label{sec:method}

To study the dynamical formation of BBHs in different galaxies, we apply the method presented in \citetalias{grudic22} and \citetalias{rodriguez23} to a suite of dwarf galaxies in addition to the MW like galaxy analysed in those papers. We use this set of cosmological simulations (\S \ref{subsec:sims}) combined with a cluster formation framework (\S \ref{subsec:CMC}) to model populations of star clusters. These clusters are then evolved forward in time (\S \ref{subsec:CMC}) to produce a collection of BBHs formed in star clusters over the course of cosmic time for each galaxy. A schematic description of our complete procedure is shown in Figure \ref{fig:cartoon}.

\begin{figure*}
    \begin{subfigure}{.315\textwidth}
    \centering
    \includegraphics{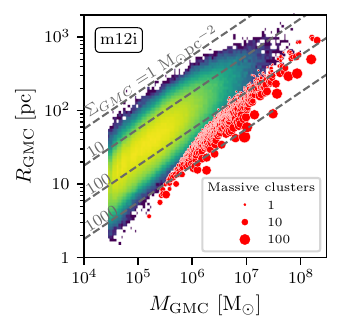}
    \end{subfigure}%
    \begin{subfigure}{.315\textwidth}
    \centering
    \includegraphics{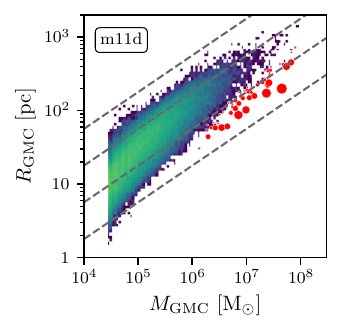}
    \end{subfigure}
    \begin{subfigure}{.365\textwidth}
    \centering
    \includegraphics{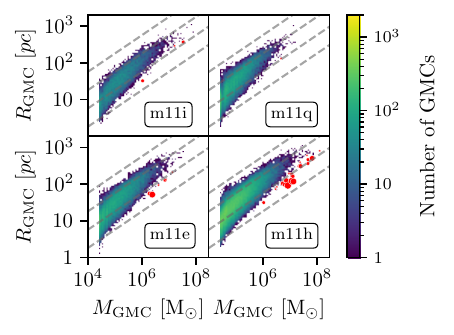}
    \end{subfigure}%
    \caption{Distributions in masses and radii of all the GMCs identified in the MW like galaxy \texttt{m12i} (\textbf{left}), in \texttt{m11d} (\textbf{center}), and in the four other cosmological simulations of dwarf galaxies considered in this study (\textbf{right}). The grey diagonal dashed lines represent lines of constant surface density $\Sigma_\mathrm{GMC}\equiv M_\mathrm{GMC}/\pi R_\mathrm{GMC}^2$. Red dots indicate the number of massive clusters ($M_\mathrm{cl}\geq6\times10^4\ \Msun$) formed in each GMC.}
    \label{fig:GMCs}
\end{figure*}

\subsection{Realistic star formation models from FIRE-2 zoom-in simulations}
\label{subsec:sims}

We make use of cosmological zoom-in simulations of dwarf galaxies described in the \textit{Core} suite of the public data release \citep{wetzel22} from the Feedback in Realistic Environment \citep[FIRE;][]{hopkins14} project. These simulations were generated with the hydrodynamics and gravity solver \texttt{GIZMO} \citep{hopkins15} using the mesh-free-finite-mass (MFM) method. Each simulation uses the second-generation FIRE-2 physics model \citep{hopkins18}, which includes gas cooling and heating, star formation, and stellar feedback. We now briefly describe the most relevant details of the FIRE-2 model.

Radiative cooling and heating is modeled in FIRE-2 across the range $10-10^{10}\ \mathrm{K}$. This includes free-free, photo-ionization and recombination, Compton, photo-electric and dust collisional, cosmic ray, molecular, metal-line, and fine-structure processes. This also contains photo-ionization and heating from a redshift-dependent, spatially uniform ultraviolet background \citep{faucher09}. 
Star formation occurs in self-gravitating, self-shielding, and Jeans unstable dense molecular gas ($n > 1000\ \mathrm{cm}^{-3}$). Stellar feedback includes energy, momentum, mass and metal injection from type Ia and type II supernovae and stellar winds (from O, B and AGB stars). All feedback event rates, luminosities, energies, and mass-loss rates are tabulated from stellar evolution models \citep[\texttt{STARBUST99;}][]{leitherer99} assuming a \cite{kroupa01} initial stellar mass function (IMF). All simulations include a sub-grid model for the turbulent diffusion of metals in gas, which produces more realistic metallicity distributions \citep{escala17}.

The first five simulations considered in this work correspond to \texttt{m11q}, \texttt{m11i}, \texttt{m11e}, \texttt{m11h}, and \texttt{m11d} presented in \citet{elBadry17}. They all have the same mass resolutions, with $m_b \sim 7100\ \Msun$ for baryonic particles and $m_{dm} \sim 39000\ \Msun$ for dark-matter (DM) particles. The minimum adaptive force softening length is 1 pc for gas cells, 4 pc for star particles and 40 pc for DM particles. All simulations assume a flat $\Lambda$CDM cosmology consistent with recent measurements \citep{planck20} $\Omega_{\Lambda}$ = 0.69, $\Omega_m$ = 0.31, $\Omega_b$ = 0.048, h = 0.68, $\sigma_8$ = 0.82, and n$_s$ = 0.97, with the exception of \texttt{m11q}, which uses the cosmology from the AGORA project \citep{kim13} $\Omega_{\Lambda}$ = 0.728, $\Omega_m$ = 0.272, $\Omega_b$ = 0.0455, h = 0.702, $\sigma_8$ = 0.807, and n$_s$ = 0.961. The halo DM masses at redshift $z=0$ range from $7.77\times10^{10}\ \Msun$ for the least massive halo (\texttt{m11i}) to $3.23\times10^{11}\ \Msun$ for the most massive one (\texttt{m11d}).
We add to this list of simulations the MW like galaxy \texttt{m12i} presented in \citetalias{grudic22}. Unlike the first five simulations of dwarf galaxies, this model includes a treatment of magneto-hydrodynamics, conduction, and viscosity. The effects of such additional physical processes upon galaxy evolution and star formation have been shown to be modest in the FIRE-2 simulations \citep[see e.g][]{su17, hopkins20}. Halo masses, stellar masses, and other properties of the simulated galaxies are listed in Table \ref{tab:sims}.

In Figure \ref{fig:sims} we show the evolution of the star formation rate (SFR) and the average metallicity of newly-formed stars across time in the six simulated galaxies. Each galaxy has a unique star formation history, with \texttt{m11q} having a very early peak of star formation around 12 Gyr ago while the most massive dwarf galaxy, \texttt{m11d}, had more recent and bursty periods of star formation around 6 Gyr ago. The MW like galaxy \texttt{m12i} has a SFR that is generally more regular over time in the last 10 Gyr.
The mean metallicity follows an overall similar trend in all galaxies. Metals formed in the most massive stars are dispersed through winds and supernovae in the interstellar medium, and the stars that form later from this enriched gas will naturally have higher metallicities. The average stellar metallicity of newly-formed stars in a given galaxy will therefore generally increase over time.
%Being more massive, \texttt{m12i} is also the most metal-rich galaxy across time.

\subsection{Cluster formation model}
\label{subsec:model}

As the resolution of the simulations is not sufficient to resolve star clusters directly, we follow the method introduced in \citetalias{grudic22} and use the population of giant molecular clouds (GMCs) present in the simulation as the formation sites of these clusters. The GMCs are identified at all redshifts as self-gravitating gas structures using the \texttt{CloudPhinder} algorithm \citep{guszejnov19}. The number of GMCs identified in a given simulation generally increases with the mass of the DM halo, with the exception of \texttt{m11d}, which is more massive than \texttt{m11h} and yet has less GMCs.

To determine the properties of the star clusters formed in a given GMC, we use the cluster formation framework introduced in \citet{grudic21}. This model was calibrated with high-resolution magneto-hydrodynamic simulations of collapsing gas clouds. Here we briefly summarize the mapping procedure from a given GMC to its star clusters. We first compute the total stellar mass formed in the cloud $\mathrm{M}_{\star}=\epsilon_{int}M_\mathrm{GMC}$, where $\epsilon_{int}$ is the integrated star formation efficiency, itself estimated from the cloud surface density $\Sigma_{\mathrm{GMC}}\equiv M_{\mathrm{GMC}}/\pi R^2_{GMC}$. We calculate the fraction of this mass locked in bound clusters $f_{\mathrm{bound}}$, which also takes the form of an increasing function of the cloud surface density $\Sigma_{\mathrm{GMC}}$, and obtain the total mass of stars in bound clusters $M_{\mathrm{bound}}=f_{\mathrm{bound}}M_{\star}$. Individual cluster masses are then sampled from a cloud-level mass distribution until their sum reaches $M_{\mathrm{bound}}$. This cloud-level mass distribution is taken to be a power-law with an exponential cutoff, where the slope of the power law and the position of the cutoff are both functions of the cloud metallicity. We finally associate each cluster with a half-light radius sampled from a cloud-level size-mass relation. The cluster metallicity is taken to be the same as its parent GMC. We refer the reader to Section 4 in \citet{grudic21} for further details on the cluster formation framework.

The simulations of gas clouds used to build this cluster formation model are presented in \citet{grudic21}. \citet{su17, hopkins20} show that the effects of additional physical processes such as MHD, conduction, or viscosity do not have a significant effect on the bulk properties of the simulated galaxy. We checked the consistency of our cluster formation model by comparing the results presented in \citetalias{grudic22} with the same analysis applied to the non-MHD version of the \texttt{m12i} galaxy \citep{wetzel16}. In this non-MHD simulation, we find GMCs that tend to be more massive, although less dense, than the clouds identified in \citetalias{grudic22}. This results in an excess of massive clusters. We expect this difference to become less important in dwarf galaxies, where the distributions naturally converge towards lower GMC masses, even in the absence of any of these additional physical processes.

It is important to note that the list of GMCs identified with \texttt{CloudPhinder} is only a subset of all the clouds that would ever exist in a similar real galaxy. Indeed, the lifetime of a GMC found in these simulations is generally in the order of $\sim 1 - 10\ \mathrm{Myr}$ \citep{hopkins12, chevance19, benincasa20}, which is significantly less than the average time interval between consecutive snapshots in our simulations ($\Delta t\sim 20 \mathrm{Myr}$). A number of gas clouds could spend their entire life cycle within a single time step of the simulation and thus never appear in the simulation output. To correct for this incompleteness of our GMC catalogue, at each snapshot we repeat the sampling procedure for our catalog of identified GMCs until the total mass of stars formed in clouds reaches the stellar mass formed in the entire galaxy between this snapshot and the next. In this way, our model accounts for all the stellar mass formed in the simulated galaxy. The mass-radius distributions of all the GMCs identified with \texttt{CloudPhinder} in our 6 simulated galaxies are shown in Figure \ref{fig:GMCs}. Red dots indicate the GMCs that form massive clusters ($M_\mathrm{cl}\geq6\times10^4\ \Msun$). The model predicts a distinct area on the mass-radius plane that is more conducive to the creation of massive clusters. This apparent preference for the formation of massive clusters in clouds of high mass and surface density is consistent with analytical expectations that stellar feedback becomes progressively inefficient as densities increase \citep{kruijssen12}. The effect of the galactic environment on the cluster formation efficiency is also supported by several observations \citep[see e.g.][]{adamo15, johnson16}.

To maintain consistency with the cluster catalog presented in \citetalias{grudic22}, we only select clusters with an initial mass greater than $10^3\ \Msun$. From this set of sampled clusters, we define the population of massive clusters as $M_\mathrm{cl} \geq 6\times 10^4\ \Msun$, which will be evolved forward in time. This threshold on the cluster initial mass is motivated both by the fact that we are focusing on the clusters most likely to host the majority of dynamically formed and hierarchical BBH mergers, and to ensure that they have sufficient particles to be integrated with a H\'enon-style Monte Carlo approach, which we describe below \citepalias[see also][Section 2]{rodriguez23}.

\subsection{Dynamical assembly in star clusters with \texttt{CMC}}
\label{subsec:CMC}

The initial positions and velocities of the individual particles in these massive clusters are generated using the Elson, Fall and Freeman (EFF) density profiles, originally developed to describe young clusters in the LMC \citep{elson87}:
\begin{equation}
    \rho(r)=\rho_0 \left(1+\frac{r^2}{a^2} \right)^{-\frac{\gamma+1}{2}} ,
\end{equation}
where $\rho_0$ is the central cluster density, $a$ is a scale radius and $\gamma$ is a power-law index drawn for each cluster according to the cluster formation model of \citet{grudic21}.

The massive clusters sampled in our zoom-in cosmological simulations are integrated forward in time using the Cluster Monte Carlo code \citep[\texttt{CMC}][]{pattabiraman13, rodriguez22}. \texttt{CMC} uses a H\'enon's Monte Carlo approach \citep{henon71} to model collisional stellar dynamics with an orbit averaging technique. It assumes that clusters with a sufficiently large number of particles ($\gtrsim 10^5$) evolve mainly by two-body encounters that can be modelled as a statistical process. This condition, known as the Fokker-Planck approximation, dictates the initial mass threshold applied to our clusters. The code includes various processes relevant to the formation of BBHs including two-body relaxation \citep{joshi00}, three-body binary formation \citep{morscher15} and direct integration of small-N resonate encounters with post-Newtonian corrections \citep{rodriguez18b}. Stars and stellar binaries are evolved using the rapid population synthesis code \texttt{COSMIC} \citep{breivik20}.

Here we summarise the main assumptions used in our version of \texttt{COSMIC}. Stellar populations are initialised with primary masses sampled from a \citet{kroupa01} IMF between 0.08 $\Msun$ and 150 $\Msun$ and uniformly distributed mass ratios. Following \citet{vanHaaften13} we assume a mass-dependent binary fraction $b(M)=\frac{1}{2}+\frac{1}{4}\mathrm{log}_{10}(M)$. As the star clusters in the \texttt{m12i} galaxy had already been evolved using CMC with a fixed binary fraction $b=0.2$ (see \citetalias{rodriguez23}), we decided not to run them again due to the high computational cost. 
%\sout{As a consistency check, we simulated some of the new clusters with these two different prescriptions and verified that there was no major statistical difference.}
\citet{chatterjee17} show that, because most initial binaries are quickly disrupted by dynamical interactions, varying the initial population of stellar binaries in a cluster does not have a strong impact on the final number and properties of BBHs retained inside the cluster. One exception to this effect is the possibility, in the case of a large number of initial massive stellar binaries inside a small cluster (with initial virial radius $r_\mathrm{v}\lesssim1$ pc), of collisional runaways leading to the formation of an intermediate mass black hole \citep[see e.g.][]{gonzalez21}.

Metallicity-dependent wind mass loss rates for O and B stars are set by \citet{vink01} and winds from Wolf-Rayet start are treated according to \citet{vink05}. We assume a value for solar metallicity of $\Zsun=0.017$ \citep{grevesse98}. The stability of mass transfer events is determined following \citet{neijssel19}, where the criteria is whether or not the donor star expands faster than its Roche-lobe radius in response to the mass transfer. An exception is made for stripped stars whose mass transfer is assumed to be always dynamically stable. Common envelope (CE) evolution follows the standard $\alpha \lambda$ model described in \citet{hurley02}, with the efficiency parameter $\alpha$ set to 1 and the binding energy factor $\lambda$ determined by \citet{claeys14}. Remnant masses are computed with the \textit{delayed} prescription from \citet{fryer12}. Natal kicks for BHs and NSs are drawn from a Maxwellian distribution with a dispersion $\sigma = 265 \mathrm{km\ s}^{-1}$ \citep{hobbs05} and are fallback-modulated in the case of BHs \citep{fryer12}. We refer the reader to \citet{breivik20} for further details on the different physical prescriptions implemented in \texttt{COSMIC}.

To fully describe the evolution of our clusters in their respective galaxies, we also need to take into account the tidal effects of the host galaxy on the clusters. Following \citetalias{rodriguez23}, each cluster is associated with a star particle selected from its parent GMC. These particles are used as tracers of realistic trajectories for our clusters in each galaxy. Without taking into account dynamical friction (discussed below), this gives us trajectories consistent with the cosmological simulation for all sampled clusters. 

The external tidal field experienced by these tracer particles is then extracted directly along the path of the tracer particle. We compute at each snapshot the tidal tensors as
\begin{equation}
    \mathbf{T}^{ij} \equiv -\left(\frac{\partial^2\phi}{\partial x^i \partial x^j} \right)_{\mathbf{r}} ,
\end{equation}
where $\phi$ is the gravitational potential evaluated at the position $\mathbf{r}$ of the tracer particle in the reference frame $\{x^i\}_{i=1,2,3}$. The tidal tensors are passed as input to \texttt{CMC} and are diagonalised to compute an \textit{effective} tidal strength as $\lambda_{1,e}\equiv \lambda_1-(\lambda_2+\lambda_3)/2$ where $\lambda_{1,2,3}$ are the eigenvalues sorted from largest to smallest \citep{pfeffer17}. This allows us to compute, at each time step and for each cluster, its instantaneous tidal radius as
\begin{equation}
    r_\mathrm{t}=\left(\frac{GM_\mathrm{c}}{\lambda_{1,e}}\right)^{1/3} .
\end{equation}
All stars with orbital apocenters greater than $r_\mathrm{t}$ are eliminated by \texttt{CMC}. 

As star clusters move through their host galaxy, they are subject to dynamical friction caused by their surroundings (stars, gas, dark matter). They lose orbital energy and angular momentum, and eventually spiral towards the galactic center. This effect is not taken into account in the orbits of the tracer particles and must be added in post-processing.
Although the interplay between tidal fields and dynamical friction is particularly difficult to model and we cannot self-consistently calculate the real orbits of our clusters, we can estimate the time it would take them to spiral into the galactic center \citep[following][]{pfeffer17}. We compute for each cluster and at each snapshot this dynamical friction timescale as \citep{lacey93}:
\begin{equation}
    T_\mathrm{df}=\frac{\epsilon^{0.78}}{2B(v_\mathrm{c}/\sqrt{2}\sigma)}\frac{\sqrt{2}\sigma r_\mathrm{circ}^2}{GM_\mathrm{c}\mathrm{ln}\Lambda} ,
\end{equation}
where $r_\mathrm{circ}$ is the radius of the circular orbit with the same energy as the real orbit, $v_\mathrm{c}$ is the circular velocity at that radius, $\sigma$ is the local velocity dispersion, $M_\mathrm{c}$ is the time-dependent cluster mass, $B(x)=\mathrm{erf}(x)-2(x/\sqrt{\pi})\exp(-x^2)$ is the standard
velocity term for dynamical friction \citep{binney08}, $\epsilon$ is the ratio of the real angular momentum to that of the circular orbit, and $\mathrm{ln}\Lambda$ is the Coulomb logarithm. Folowing \citet{lacey93}, $\Lambda$ is defined as $\Lambda=1+M_\mathrm{c}/M_\mathrm{enc}$ where $M_\mathrm{enc}$ is the mass enclosed.

Each cluster is integrated forward in time until the condition 
\begin{equation}
    \int \frac{dt}{T_\mathrm{df}(t)} > 1
\end{equation}
is satisfied. From this point onward, the cluster is assumed to have merged with the galactic center.

We refer the reader to \citetalias{grudic22, rodriguez23} for more information on the population of star clusters sampled in the MW like galaxy \texttt{m12i}.

\section{Results}\label{sec:results}

In this section, we present the outcomes of our cluster formation framework applied to our set of zoom-in cosmological simulations and some properties of the BBHs dynamically formed therein. Firstly, we examine the characteristics of star clusters in all simulated galaxies (\S \ref{subsec:clusters}) and their efficiency in producing BBHs (\S \ref{subsec:eta}). We then study the physical properties of these BBHs and investigate the relationship between the BBH mergers and the star formation history exclusive to each galaxy (\S \ref{subsec:bbhs} and \S \ref{subsec:environment}).
 
\subsection{Cluster physical properties}
\label{subsec:clusters}

\begin{figure}
    \centering
    \includegraphics{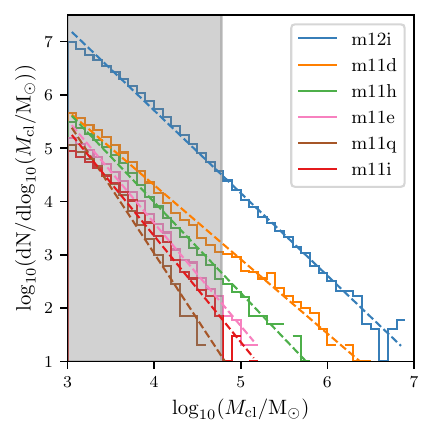}
    \caption{Initial mass distributions of all clusters sampled in our six zoom-in simulations of galaxies. We consider here all the clusters formed without taking into account their different redshifts of formation. The different simulations are indicated with the same colors as in Figure \protect{\ref{fig:sims}}. Dashed lines correspond to the best power-law fits.
    The grey shaded region indicates the mass range $M_\mathrm{cl}<6\times10^4\ \Msun$ that we do not model with \texttt{CMC}. This limit roughly corresponds to an initial number of particles of $10^5$.}
    \label{fig:gcmf}
\end{figure}

\begin{table}
    \centering
    \begin{tabular}{c|c|c|c}
    %\hline
        \multirow{2}{*}{Simulation} & Total mass in clusters & \multirow{2}{*}{Massive clusters} & \multirow{2}{*}{$\alpha$}\\
         &  [$\Msun$] & &  \\
        \hline
        m11i & $8.65\times10^7$ & 5 & -2.99 \\
        m11q & $7.96\times10^7$ & 1 & -3.48 \\
        m11e & $1.43\times10^8$ & 20 & -2.93 \\
        m11h & $3.09\times10^8$ & 97 & -2.70 \\
        m11d & $5.72\times10^8$ & 482 & -2.39 \\
        m12i & $1.37\times10^{10}$ & 8575 & -2.55 \\
    %\hline
    \end{tabular}
    \caption{Properties of the clusters sampled in our six zoom-in simulations of galaxies. The selection criterion of massive clusters is $M_\mathrm{cl}\geq6\times10^4\ \Msun$. It is only applied at the epoch the cluster is formed. The $\alpha$ parameter corresponds to the best fit power-law index described in Eq. \protect{\ref{eq:powerlaw}}.}
    \label{tab:slopes}
\end{table}

The initial mass distributions of all the clusters sampled in our six zoom-in simulations are shown in Figure \ref{fig:gcmf}. We fit the mass distributions with power-law functions
\begin{equation}
    \Phi(M) \equiv \frac{\mathrm{d}N}{\mathrm{d}M} \propto M^{\alpha} ,
    \label{eq:powerlaw}
\end{equation}
and show the values obtained for $\alpha$ in Table \ref{tab:slopes}. In addition to having a larger total number of star clusters, the more massive galaxies \texttt{m11d} and \texttt{m12i} also clearly extend to higher masses, while having shallower power-law slopes overall.
The steep slopes of the cluster initial mass functions in our smallest dwarf galaxies, along with the expectation that the total mass of globular clusters in a galaxy relates to its stellar mass \citep[see e.g.][]{beasley20, berek23}, suggests that it is highly improbable to find massive clusters in low-mass galaxies. 
This result is in good agreement with observations of star clusters in the Local Group \citep[see e.g. Figure 10 in][for a comparison of the mass functions of young clusters in different galaxies]{portegies10}. 
% If the GW signal from a BBH observed with the LVK interferometers allows us to determine that it formed in a massive cluster, it would then be possible to add constraints on the mass of the galaxy in which it formed.

\begin{figure}
    \centering
    \includegraphics{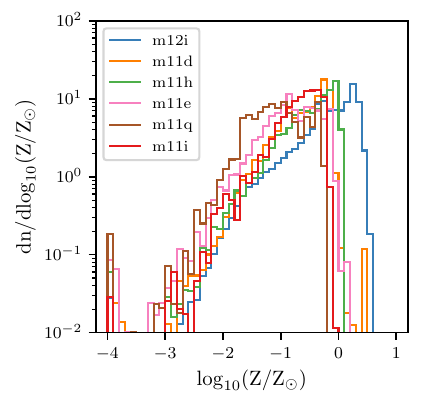}\\
    \includegraphics{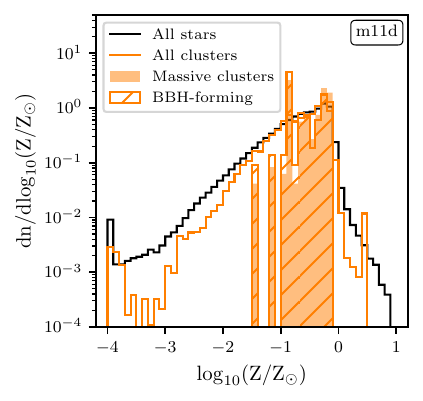}\\
    \includegraphics{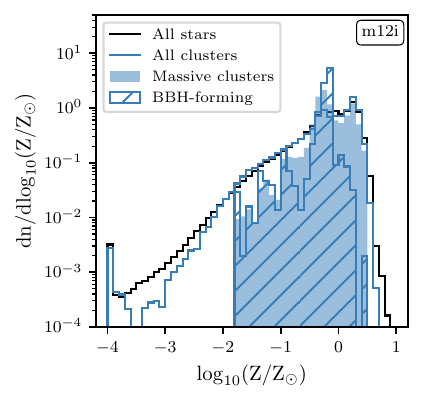}
    \caption{Metallicity distributions of all clusters sampled in our six zoom-in simulations of galaxies (\textbf{Top}). The different simulations are indicated with the same colors as in Figure \protect{\ref{fig:sims}}. The minimal value of metallicity admitted in the FIRE-2 simulations is $\sim10^{-4}\Zsun$, which causes a stacking at low-metallicity visible on the plot.\\ %It has been shown that this unnatural artefact has no effect on the various results presented in this paper.
    Metallicity distributions in the simulations \texttt{m11d} (\textbf{Center}) and \texttt{m12i} (\textbf{Bottom}) of all stars (black line), all clusters (coloured line), massive clusters (filled area), and clusters in which merging BBHs form (hatched area). %In theses two subplots, we normalised all the histograms to one.
    All histograms are normalised to one.
    %to make the comparison between different populations easier to read.
    }
    \label{fig:clusterZ}
\end{figure}

In the top panel of Figure \ref{fig:clusterZ} we show the distributions of cluster mean metallicities at birth in our six cosmological simulations. In our cluster formation framework, the metallicity of a cluster is directly inherited from its parent GMC. In all six simulated galaxies the distributions exhibit a similar trend, with a metallicity peak between $0.1$ and $1\ \Zsun$ depending on the galaxy considered, and a sharp drop just after this peak.
For each galaxy in our collection, the cluster metallicity distribution has a shape almost identical to that of the complete population of star particles in the simulation (see the examples of \texttt{m11d} and \texttt{m12i} in the other panels). 

Looking individually at different cluster populations in \texttt{m11d} (center panel of Figure \ref{fig:clusterZ}), we note that there is a preference for the formation of massive clusters within a fairly narrow metallicity range compared to the distribution of all clusters. The lower limit comes naturally from the fact that metallicity is lowest at early times, when star formation has only just begun (see Figure \ref{fig:sims}). At that stage, dwarf galaxies are even smaller and do not contain GMCs massive enough to populate the high-mass tail of the cluster mass function. 
On the other hand, the upper limit of the massive cluster metallicity distribution function is expected around the maximum mean metallicity that the galaxy reaches, and so is expected to track the usual galaxy mass-metallicity relation \citep{kruijssen19}.
The massive clusters in galaxy \texttt{m12i} cover a wider range of metallicities (bottom panel of Figure \ref{fig:clusterZ}). This is because the simulated galaxy gets more massive at earlier times compared to the dwarf galaxies, and therefore contains sufficiently massive and dense GMCs when the metallicity is still low. Massive clusters formed at high metallicities, not present in dwarf galaxies, come from massive, dense and metal-rich GMCs that exist only in our MW like galaxy \texttt{m12i}.

We note however that, as described in \citetalias{grudic22, rodriguez23}, our massive star clusters sampled in \texttt{m12i} are significantly younger and more metal-rich than the GCs observed in the MW and in M31 \citep[see e.g.][for ages and metallicities of GCs in the MW and M31 respectively]{harris96, caldwell11}. \citetalias{rodriguez23} suggested that this discrepancy could be explained by differences between the star formation history of the MW galaxy and that of the simulated galaxy \texttt{m12i}. The specific star formation histories of our cosmological simulations and the correlations with BBH mergers will be studied in more detail below in \S \ref{subsec:bbhs}.

\begin{figure*}
    \begin{subfigure}{.315\textwidth}
    \centering
    \includegraphics{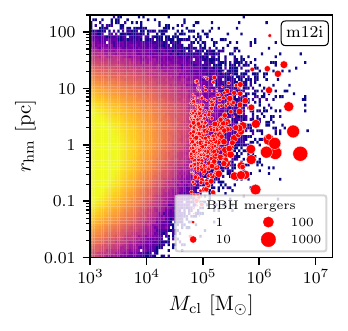}
    \end{subfigure}%
    \begin{subfigure}{.315\textwidth}
    \centering
    \includegraphics{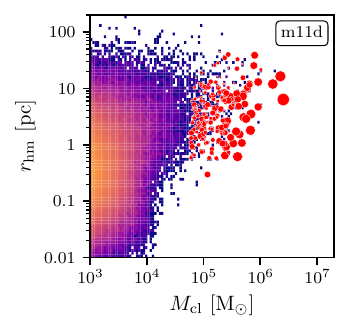}
    \end{subfigure}
    \begin{subfigure}{.365\textwidth}
    \centering
    \includegraphics{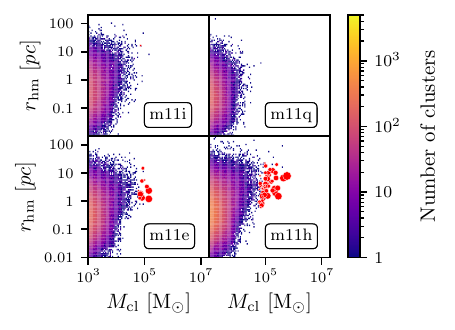}
    \end{subfigure}%
    \caption{Masses and radii of all the clusters sampled in the MW like galaxy \texttt{m12i} (\textbf{left}), in \texttt{m11d} (\textbf{center}), and in the four other cosmological simulations of dwarf galaxies considered in this study (\textbf{right}). Red dots indicate the number of BBHs formed in each cluster.}
    \label{fig:clusters}
\end{figure*}

\subsection{Formation of BBHs in star clusters}
\label{subsec:eta}

Here we look specifically at the star clusters in which BBHs are formed, and we describe how populations of BBHs arise from various groups of clusters in different galaxies.
In Figure \ref{fig:clusters} we show the mass-radius distribution of the star clusters sampled in our six zoom-in simulations. The clusters in which BBHs formed are pictured as red dots, where the size of the dot indicates the number of BBHs. In the case of the MW like galaxy \texttt{m12i}, only $\sim1/3$ of the massive clusters were evolved with \texttt{CMC} to reduce the computational cost. This random sampling is the reason for the apparent lack of clusters forming BBHs compared with the complete population in the left plot. 

There appears to be a clear preference for BBH formation in massive and compact clusters. More massive clusters are naturally larger reservoirs of massive stars, and are therefore more likely to host BBHs. In addition, the dynamical interactions in these dense environments tend to harden binaries that are already hard (the so-called 'Heggie's law'), thereby increasing the fraction of BBHs that merge within an Hubble time \citep[see e.g.][for more details on the relationship between the physical properties of a cluster and the population of BBHs it produces]{rodriguez16, kremer20}.

% In Figure \ref{fig:wealth} we show the distributions of the number of BBH mergers as a function of cluster initial mass in the three most massive dwarf galaxies \texttt{m11e}, \texttt{m11h}, and \texttt{m11d}, as well as in the MW like galaxy \texttt{m12i}.
As there are no clusters more massive than $1.5\times10^5\ \Msun$ in \texttt{m11e}, all BBHs are formed in clusters with an initial mass around $\sim10^5\ \Msun$. On the other hand, for \texttt{m11h} and \texttt{m11d}, which both have clusters as massive as $5\times 10^5\ \Msun$, the contribution of the most massive clusters to the overall population of BBH mergers compares with that of the less massive ones. These results suggest that a small number of very massive clusters that are very efficient at producing BBHs can make up for a large number of less massive clusters. The MW like galaxy \texttt{m12i} presents an extreme case of this effect, with one single cluster of mass $\sim 7\times 10^6 \Msun$ accounting for more than $1/4$ of all the BBHs formed from all of the 495 clusters simulated from that galaxy \citep[see][for a specific study of this 'Behemoth' cluster]{rodriguez20}.

Metallicity is one of the elements often described as playing a major role in the formation of BBHs. Indeed, as mass loss through stellar winds increases with metallicity, massive stars formed in metal-rich environments lose a significant fraction of their mass before the end of their lives. In the case of isolated binaries, this effect produces a widening of the orbits (due to the conservation of angular momentum) and generally reduces the fraction of BBHs that can merge within an Hubble time \citep{dominik12, giacobbo18, neijssel19}. In contrast, metallicity is expected to have a weaker effect on the BBH formation efficiency in star clusters, where dynamical interactions play a dominant role, and to contribute mainly to shaping the BBH mass distribution \citep[see e.g.][]{diCarlo20, kremer20}.
The two bottom subplots in Figure \ref{fig:clusterZ} show the metallicity distribution of all the massive clusters (filled) and those hosting BBH mergers (hatched) in the two simulated galaxy \texttt{m11d} and \texttt{m12i}. In \texttt{m11d}, the subset of massive clusters in which BBHs form has a metallicity distribution almost identical to that of all massive clusters. This is consistent with the idea that metallicity has a modest impact on the BBH formation efficiency in massive clusters, for sub-solar metallicity values. This effect becomes significant for the most metal-rich massive clusters in \texttt{m12i}, where the formation efficiency of merging BBHs drops rapidly with increasing metallicity above $\Zsun$. It is also important to bear in mind that these metal-rich clusters generally form late in the life of their host galaxy and that, as a result, only a small subset of all the BBHs formed within them will have coalesced before the present-day.

\begin{figure*}
    \centering
    \includegraphics{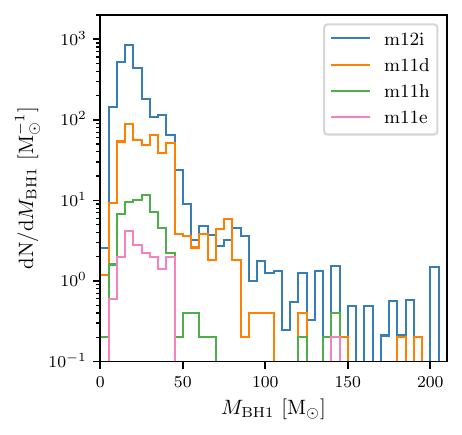}
    \includegraphics{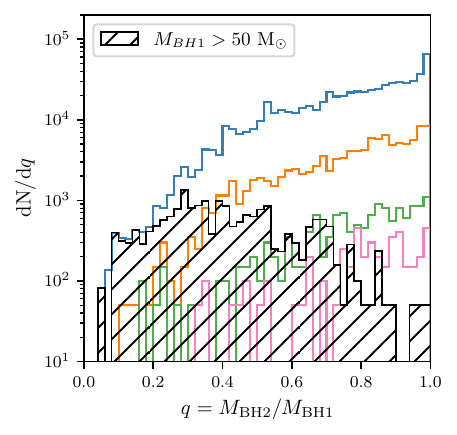}
    \caption{BBH primary mass (\textbf{left}) and mass ratio (\textbf{right}) distributions in the simulated galaxies \texttt{m11e}, \texttt{m11h}, and \texttt{m11d}. On the right-hand side plot, the hatched areas show the mass ratio distributions for the BBHs with a primary mass $M_{BH1}>50\ \Msun$ in all four galaxies combined.}
    \label{fig:mBH1&q}
\end{figure*}

\begin{figure}
    \centering
    \includegraphics{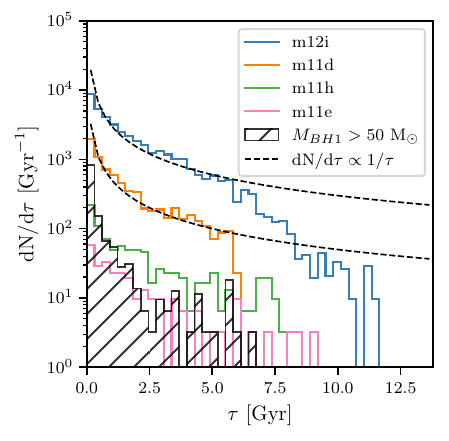}
    \caption{Distributions of delay times for all BBHs in the simulated galaxies \texttt{m11e}, \texttt{m11h}, \texttt{m11d}, and \texttt{m12i}. The hatched areas show the distribution of delay times for the massive BBHs ($M_\mathrm{BH1}>50\ \Msun$) in all four galaxies combined. The dashed black lines show $1/\tau$ distributions matching the delay times in \texttt{m11d} and \texttt{m12i}.}
    \label{fig:delays}
\end{figure}

\subsection{Properties of dynamically formed BBHs}
\label{subsec:bbhs}

Here we describe the properties of the BBHs formed in the star clusters sampled in our four cosmological simulations of galaxies \texttt{m11e}, \texttt{m11h}, \texttt{m11d}, and \texttt{m12i}. We will also examine the interplay between the star formation history and the BBH merger rate in each galaxy. As the two smallest dwarf galaxies \texttt{m11q} and \texttt{m11i} have only 1 and 5 massive clusters respectively, the number of BBHs formed in them is too small to be statistically significant. Therefore, we have chosen not to include them in the following analysis.

The distributions of primary mass $M_\mathrm{BH1}$ and mass ratio $q \equiv M_\mathrm{BH2}/M_\mathrm{BH1}$ for the BBHs formed in \texttt{m11e}, \texttt{m11h}, \texttt{m11d} and \texttt{m12i} are shown in Figure \ref{fig:mBH1&q}. 
The primary masses of BBHs display a similar distribution across the four simulated galaxies, with a peak at around $\sim 20\ \Msun$ and a sharp decrease after $\sim 45\ \Msun$.
% There is a dearth of low-mass BHs ($M_\mathrm{BH1}\leq 10\ \Msun$), a peak around $\sim 20\ \Msun$ and a sharp decrease after $\sim 45\ \Msun$. 
This drop comes from the pair-instability supernova (PISN) and pulsational pair-instability supernova (PPISN) mechanisms, in which the production of electron–positron pairs in the core of a very massive star is expected to suppress photon pressure and cause the core to contract. For stars with helium cores more massive than $m_\mathrm{He}\gtrsim60\Msun$, the collapse triggers the ignition of explosive oxygen burning which completely disrupts the star and leaves no remnant behind \citep{kasen11}. For less massive stars with helium core in the range $60\gtrsim m_\mathrm{He}\gtrsim30\Msun$, the energy released by oxygen burning is not sufficient to completely disrupt the star; instead mass gets ejected in a series of energetic pulses before the star finally collapses to a BH \citep{woosley17}. In this scenario, an upper limit for the mass of BHs formed from single star evolution naturally arises. We use the prescriptions for PISN and PPISN presented in \citet{marchant19}, which predict a mass limit of $\sim45\ \Msun$, hence the drop in the primary mass distribution of BBHs in Figure \ref{fig:mBH1&q}. The existence in our simulations of BBHs with components more massive than this limit is a direct indication of dynamical interactions, such as stellar mergers \citep{kremer20} or second generation mergers \citep{rodriguez22}.

The right panel of Figure \ref{fig:mBH1&q} indicates that in all the simulated galaxies, our model predicts a distribution weighted toward equal-mass BBH mergers. The fraction of BBHs with a mass ratio $q>0.5$ consistently stands at around 90\% (92\%, 91\%, 90\% and 90\% in \texttt{m11e}, \texttt{m11h}, \texttt{m11d} and \texttt{m12i}, respectively). However, when we specifically look at at the sub-population of massive BBHs ($M_\mathrm{BH1}>50\ \Msun$), we observe a preference for unequal-mass mergers. Among all the massive BBH mergers in the four considered galaxies, 68\% have a mass ratio $q\leq0.5$, indicating that the primary BH is at least twice as massive as the secondary. Such values of mass ratios are consistent with hierarchical mergers, where a massive second generation BH (born of an earlier BBH merger) pairs with a first-generation BH (born of a massive star).

High and unequal component masses have been proposed as a signature to identify dynamically formed BBHs in GW observations \citep[see e.g.][]{sedda20}. As discussed above, the PISN and PPISN mechanisms would make it impossible to form BHs more massive than $\sim45\Msun$ through single star evolution alone. In the evolution of stellar binaries, episodes of mass transfer from an evolved star onto an already-formed BH could allow this limit to be exceeded. However, assuming that accretion onto a BH is Eddington-limited, the increase in mass of the first-born BH alone cannot explain the most massive BBH mergers reported by the LVK Collaboration \citep{bouffanais21b, abbott20}. On the other hand, super-Eddington accretion would cause these mass transfer episodes to be more conservative, thus forming BBHs that are too wide to merge within an Hubble time \citep{vanSon20, bavera21}. The existence of BHs in this mass gap cannot therefore be explained by the evolution of stellar binaries alone, and could be a direct indication of hierarchical mergers. \citet{rodriguez20} explored the formation of unequal-mass BBHs in dynamical environments and showed that the LVK event GW190412, a BBH reported with a mass ratio of $q\sim0.25$ \citep{gw190412}, is consistent with a third-generation merger in a massive, dense star cluster with a high escape velocity (which is included in our \texttt{m12i} population).

In Figure \ref{fig:delays} we show the distribution of delay times for the BBHs formed in star clusters within the four same simulated galaxies. Here the delay time refers to the time elapsed from the formation of the cluster to the BBH merger. The starting point is always set at the formation of the cluster, whether the BBH is a first-generation merger or not. This definition allows us to relate BBH mergers directly to the galactic environment in which they were formed, via their parent star cluster. In the case of observed GW events small delay times would indicate that the BBHs merged shortly after the creation of their star clusters, which therefore could still be observable as young clusters in the local Universe. 
The distributions of delay times in the four galaxies have an overall similar shape, consistent with a log-uniform distribution (or $\mathrm{dN}/\mathrm{d}\tau \propto 1/\tau$) \citep[see e.g.][]{dominik12, rodriguez18, neijssel19}. Short delay times are therefore strongly favoured, with a majority of BBH mergers taking place less than 2 Gyr after the formation of their parent cluster (68\%, 63\%, 74\% and 67\% in \texttt{m11e}, \texttt{m11h}, \texttt{m11d} and \texttt{m12i}, respectively). 
%The most rapid mergers take place around $\sim10$ Myr after cluster formation in all galaxies.
The sub-population of massive BBH mergers ($M_\mathrm{BH1}>50\ \Msun$) is even more prone to short delay times, with 86\% of them having a delay time $\tau<1\ \mathrm{Gyr}$ in all four galaxies combined. 
We note that there appears to be a scarcity of very long delay times ($\tau>6$ Gyr) in the simulated galaxies \texttt{m12i} and \texttt{m11d} compared to a $1/\tau$ distribution function. This characteristic is due to the fact that, in our simulations, most clusters are formed relatively late and therefore do not allow for long delay times before merger. Except for dynamical friction, allowing these clusters to evolve for longer times would result in a smooth power law. We explore the specific star formation histories and cluster formation histories of our simulations in the next section.

\begin{figure*}
    \centering
    \includegraphics{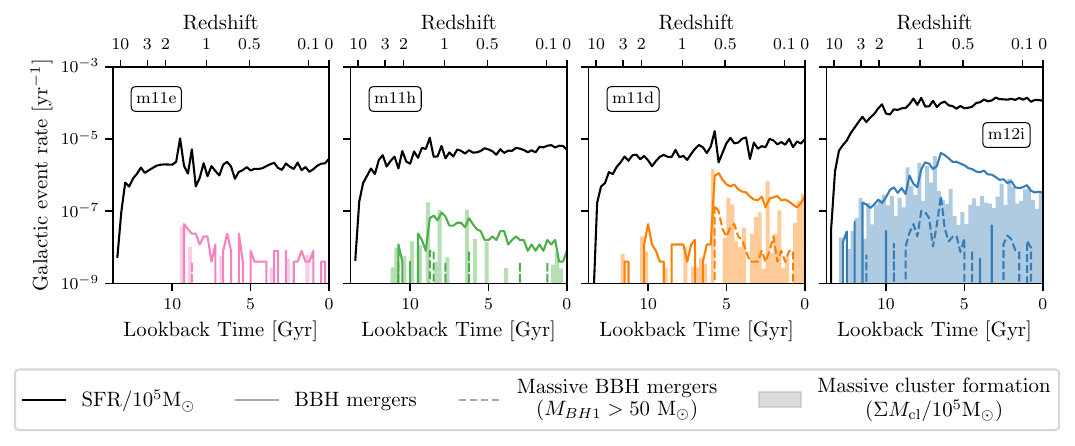}
    \caption{Dynamical BBH merger rate in the simulated galaxies \texttt{m11e}, \texttt{m11h}, \texttt{m11d}, and \texttt{m12i} (left to right). In each subplot, the BBH merger rate is shown as a coloured solid line (number of mergers per unit time). Dotted lines represent the BBH merger rates of massive BBHs ($M_\mathrm{BH1}>50\ \Msun$).
    The filled areas correspond to the formation rate of massive clusters. They indicate the total mass of massive clusters formed, divided by $10^{5}\ \Msun$, per unit time. 
    The black solid line shows the SFR, divided by $10^{5}\ \Msun$, for comparison.}
    \label{fig:merger_rates}
\end{figure*}

\subsection{BBHs progenitor environment}
\label{subsec:environment}

Combining the epochs of formation of the clusters sampled in our simulations and the delay times of the BBHs formed in each of these clusters, we can now populate our galaxies with BBH mergers across cosmic time. Figure \ref{fig:merger_rates} shows the evolution of the massive cluster formation rate and of the BBH merger rate $\mathcal{R}_{\mathrm{BBH}}$ as a function of time. Most massive cluster formation events (shown as filled areas) are closely followed by a sharp increase in the BBH merger rate. This pattern arises from the preference for short delay times described above.
We also compare the BBH merger rates obtained with the SFR in each galaxy. The peaks in the BBH merger rate are sparse and correlate to only a few of the highest star formation peaks, as the massive clusters themselves form preferentially from the most massive and densest molecular clouds.

In our simulations, we observe a positive correlation between the formation of massive clusters, SFR peaks and major halo merger events. We used the halo finder \texttt{ROCKSTAR} \citep{behroozi13} to identify halos in each simulation, and we built merger trees using \texttt{ytree} \citep{ytree}. For example, the peak of star formation at redshift $z\sim1.5$ ($\sim9.5$ Gyr ago) in the galaxy \texttt{m11e} corresponds to a halo merger with a mass ratio $R_\mathrm{m}\equiv \mathcal{M}_{\star,2} /\mathcal{M}_{\star,1}\sim 0.76$, where $\mathcal{M}_{\star,1}$ is the stellar mass of the main halo before the merger and $\mathcal{M}_{\star,2}$ that of the secondary halo. This event also relates to the first formation of massive clusters and is apparent in the BBH merger rate. In \texttt{m11d}, a halo merger $R_\mathrm{m}\sim 0.72$ at redshift $z\sim 0.6$ triggers a starburst episode with the SFR peaking at around $2\ \Msun \mathrm{yr}^{-1}$. The consecutive formation of massive clusters generates an increase in the BBH merger rate by almost 2 orders of magnitude.
The correlations observed here are in good agreement with recent studies showing that bursts of star formation associated with major halo mergers favour the formation of massive clusters \citep[see e.g.][]{li17}.

\section{Discussion}\label{sec:discussion}

\subsection{Caveats}
\label{subsec:caveats}

With different studies based on population synthesis methods predicting a wide range of BBH merger rates for different formation channels \citep[see e.g.][]{wong21, zevin21, santoliquido21, mapelli22, srinivasan23}, caution should be used when interpreting results from such simulations. 
However, our focus here is not on an astrophysical merger rate, but on the global connection between the physical properties of a galaxy, the massive star clusters it hosts, and the dynamical BBH mergers that form in them.

One caveat to our cluster formation framework is that, with the structure finder algorithm \texttt{CloudPhinder}, we only identify GMCs with virial parameters $\alpha_\mathrm{vir}<2$. In reality, a number of stars could form in gas clouds with virtually any virial parameter. Still, as the efficiency of star formation decreases rapidly as the virial parameter increases \citep{2012ApJ...759L..27P, kim21}, we expect the overall stellar population to be dominated by stars formed in bound gas clouds. Therefore, the fraction of stars formed in bound clusters $f_\mathrm{bound}$ used in this model should be a fairly good proxy for sampling the total mass of stars that would have formed in clusters, if the simulations had sufficiently fine resolution.

As the most massive galaxies at present-day have had a greater reservoir of stars over cosmic time, they can potentially host a greater number of star clusters. This general trend is clearly visible in the total mass of star clusters sampled in each galaxy, which increases monotonically with the present-day halo mass (see Table \ref{tab:slopes}). However, it is clear that the evolution of the star cluster formation rate across time in a given galaxy cannot be inferred by its present-day properties alone, but depends on its unique star formation history and merger history. As an example, it is striking to see the difference between the merger rate of BBHs formed in massive clusters in our galaxies \texttt{m11h} and \texttt{m11d} (\texttt{m11h} hosting 8 BBH mergers in the last Gyr while \texttt{m11d} has 177, see Figure \ref{fig:merger_rates}), even though their stellar masses at redshift $z=0$ are very similar ($3.6\times 10^9$ and $3.9\times 10^9\ \Msun$ respectively). For this reason, we believe that any model predicting the population of star clusters in a galaxy with certain present-day properties should also explore the diversity of star formation histories that can lead to these exact properties.

% Certain trends are still emerging in this study, such as the slope of the cluster mass function becoming steeper in low-mass dwarf galaxies.
A second caveat to our study of dynamically formed BBHs is that, with the approach described here, we are not able to simulate low-mass and intermediate-mass clusters. We present a first approximation of their contribution to the total number of BBH mergers in each of our simulated galaxies in \S\ref{subsec:lowmass}.

\subsection{Comparison with other studies}
\label{subsec:comparison}

Several studies have looked at the formation of stellar clusters in cosmological simulations of galaxies. In their series of papers, \citet{li17, li18, li19} simulated a suite of MW like galaxies and could model the formation of individual GMCs. Cluster formation was in turn based on a process of accretion and feedback with subgrid physics. \citet{li17} found correlations between major galaxy mergers and the main episodes of cluster formation, in good agreement with our results on the impact of each galaxy's star formation history on the formation of massive clusters (\S \ref{subsec:environment}). \citet{li18} found rather steep cluster initial mass functions with slopes between $-2$ and $-3$, consistent with our values (Table \ref{tab:slopes}).

The E-MOSAICS project \citep{pfeffer17} aimed at modeling the formation, evolution, and disruption of star clusters in the galaxies present in the EAGLE simulation \citep{schaye14}. Cluster formation is here again a subgrid process. Each star particle born in the simulation triggers the formation of a population of star clusters with a fraction of stars in bound clusters depending on the properties of the interstellar medium \citep{kruijssen12}. These clusters evolve across time in the simulations accounting for stellar mass-loss, two-body interactions, and tidal shocks. The main difference with our MW like zoom-in simulation \texttt{m12i} is that they were able to reproduce with reasonable agreement the population of GCs in the MW. More specifically, they found galaxies in their collection of simulations where massive clusters form at very early times, thus recovering the population of old low-metallicity GCs that do not appear in \texttt{m12i}.

\subsection{Local BBH mergers}
\label{subsec:local}

Using all the BBH mergers reported by the LVK Collaboration after the O3 run, \cite{gwtc3_pop} found indications of a BBH merger rate increasing with redshift, as $\mathcal{R}_\mathrm{BBH}(z)\propto (1+z)^{\kappa}$ with $\kappa=2.9^{+1.7}_{-1.8}$, and computed this rate at a fiducial redshift of $z=0.2$ to be $\mathcal{R}_\mathrm{BBH}(z=0.2)=17.9-44\ \mathrm{Gpc}^{-3}\mathrm{yr}^{-1}$. As our study set contains only 6 cosmological simulations of galaxies, the inherent stochasticity of each simulation and the importance of each galaxy's unique star formation history on BBH mergers do not allow us to construct a realistic population of BBH mergers on the scale of the local Universe. However, it is possible to compute a rough estimate of the local BBH merger rate from massive star clusters predicted by our simulations

We now consider the sub-population of BBHs that merge at redshift $z<0.2$ in our simulations, as this is the fiducial redshift at which the BBH merger rate is computed in \citet{gwtc3_pop}. In the following, this sub-population is referred to as local mergers. To build an astrophysical population from these local mergers, each simulated galaxy is associated with a number density calculated from a galaxy stellar mass function (SMF). We model the SMF with a double Schechter function \citep{schechter76}:
\begin{equation}
    \Phi(\Mstar)\mathrm{d}\Mstar = e^{-\Mstar/\mathcal{M}^{*}}\left[\Phi^{*}_1\left(\frac{\Mstar}{\mathcal{M}^{*}} \right)^{\alpha_1} + \Phi^{*}_2\left(\frac{\Mstar}{\mathcal{M}^{*}} \right)^{\alpha_2} \right]\frac{\mathrm{d}\Mstar}{\mathcal{M}^{*}} ,
\end{equation}
where the characteristic mass $\mathcal{M}^{*}$, the normalisation parameters $\Phi^{*}_{1,2}$ and the power law slopes $\alpha_{1,2}$ are taken from the \citet{tomczak14} best-fit double-Schechter function of all galaxies at redshift $0.2<z<0.5$. This number density will be used as a weighting function to take account of the mass distribution of galaxies observed in our local Universe.

\begin{table}
    \centering
    \begin{tabular}{c|c|c|c}
    % \hline
        \multirow{2}{*}{Simulation} & Local mergers & $\Phi(\Mstar)\mathrm{d}\Mstar$ & $\mathcal{R}_\mathrm{BBH}(z<0.2)$ \\
         & ($z<0.2$) & [$10^6\mathrm{Gpc}^{-3}$] & [$\mathrm{Gpc}^{-3}\mathrm{yr}^{-1}$] \\
        \hline
        m11i & 0 & \multirow{2}{*}{16.35} & \multirow{2}{*}{0} \\
        m11q & 0 & & \\
        \hline
        m11e & 10 & 10.19 & 0.04 \\
        \hline
        m11h & 25 & \multirow{2}{*}{7.70} & \multirow{2}{*}{0.82} \\
        m11d & 511 & & \\
        \hline
        m12i & 1118 & 3.05 & 1.40 \\
        \hline
        \hline
        Total & 1664 & - & 2.26
    \end{tabular}
    \caption{Local BBH mergers, number density of galaxies and local BBH merger rate for our six simulated galaxies. The galaxies are grouped in four mass bins: $8.5\leq \mathrm{log}(\Mstar/\Msun)\leq 9$, $9\leq \mathrm{log}(\Mstar/\Msun)\leq 9.5$, $9.5\leq \mathrm{log}(\Mstar/\Msun)\leq 10$, and $10.5\leq \mathrm{log}(\Mstar/\Msun)\leq 11$.}
    \label{tab:local}
\end{table}

To reduce the stochasticity inherent to our limited sample of cosmological simulations, we group the galaxies into logarithmic mass bins. We count the average number of local BBH mergers in each mass interval and combine this with the number density of galaxies in this particular mass range to obtain a local BBH merger rate. In Table \ref{tab:local} we present the results obtained at each stage of the process. By summing over all the mass bins, we finally get an estimate of the local BBH merger rate from dwarf galaxies $\mathcal{R}_\mathrm{BBH}(z<0.2)\sim 2.26\ \mathrm{Gpc}^{-3}\mathrm{yr}^{-1}$. This result falls below the BBH merger rate inferred with LVK observations $17.9-44\ \mathrm{Gpc}^{-3}\mathrm{yr}^{-1}$ \citep{gwtc3_pop}. However, it is important to note that we have only examined galaxies in a limited ($8.5\leq \mathrm{log}(\Mstar/\Msun)\leq 11$) and incomplete mass range (no simulation of galaxy in $10\leq \mathrm{log}(\Mstar/\Msun)\leq 10.5$). We have not explored the contribution of the smallest dwarf galaxies, which are not expected to host any massive clusters, or that of a galaxy more massive than the MW, which would certainly have a very large number of dynamically formed BBH mergers. In addition, this model takes into account BBHs formed in massive star clusters exclusively. The omitted contribution of lower mass clusters will be discussed in the next section below. 
%In the future, we plan to expend this study on the formation of BBHs in cosmological simulations to a more complete set of galaxies, and including less massive clusters as well as BBHs formed via the isolated field channel.

\subsection{BBH formation in low-mass clusters}
\label{subsec:lowmass}

As the code for cluster evolution \texttt{CMC} that we use is based on a H\'enon-style Monte Carlo approach (see \S \ref{subsec:CMC}), a minimum initial number of particles ($\sim10^5$) is required to ensure that the cluster relaxation timescale is significantly longer than the dynamical timescale. This corresponds to a cluster initial mass $M_\mathrm{cl}\sim6\times10^4\Msun$. We have therefore not considered in this work the formation of BBHs in less massive clusters. However, several studies have shown that young stellar clusters could make a great contribution to the population of merging BBHs. \citet{mapelli22} predict that the local BBH merger rate density $\mathcal{R}_\mathrm{BBH}(z<0.2)$ is of the same order of magnitude for BBHs formed in GCs and in YSCs in all their models.

Adopting the BBH merger efficiency as a function of metallicity presented in \citet{diCarlo20} for YSCs with masses between $10^3$ and $3\times10^4 \Msun$, we can estimate the number of BBH mergers formed within star clusters in this mass range in our cosmological simulations. We find 1078, 1307, 2351, and 32790 BBH mergers from the YSCs in \texttt{m11e}, \texttt{m11h}, \texttt{m11d}, and \texttt{m12i} respectively. These figures are comparable to and higher than the number of BBH mergers found in the most massive clusters. This other formation channel should also be taken into account to create a more complete population of BBHs in our simulations.

However, the BBHs formed in YSCs are expected to have quite different physical properties than the ones formed in GCs. \cite{torniamenti22} show that low-mass clusters form, for the most part, low-mass BBHs are shaped primarily by binary evolution. In addition, the vast majority of BBHs formed in YSCs reach coalescence after being ejected from their parent cluster. This is because the escape velocity is generally lower in these low-mass clusters. It is therefore impossible for hierarchical mergers to take place in such environments \citep{gerosa19}, which makes this a distinctive feature in the properties of the BBHs formed between these two populations of star clusters.

As most stars form in YSCs, and in particular massive stars \citep{portegies10}, it would be complicated to differentiate BBH mergers formed in low-mass clusters from isolated field mergers in our cosmological simulations, but this question is outside the scope of this work.
%We aim to address this question in a following study.

\section{Conclusions}
\label{sec:conclusions}

In this work we have modeled the formation of star clusters in six cosmological zoom-in simulations of galaxies taken from the FIRE-2 project using a cloud-to-cluster formation framework applied to the GMCs identified in each simulation. Extending on the work presented in \citetalias{grudic22, rodriguez23}, we have evolved the massive clusters with the Monte Carlo code \texttt{CMC} and studied the characteristics of the BBH mergers occurring in all six galaxies. The cosmological simulations provide self-consistent environments for star formation while the impact of the galactic environment on cluster evolution is taken into account through time-dependent tidal fields and dynamical friction. We have found distinct features that could prove decisive in better understanding the properties of BBH mergers observed so far in the first 3 observation runs and currently during the 4th LVK observing run.

\begin{enumerate}
\item The total mass of stars formed in bound clusters scales with the present-day stellar mass of the galaxy. Specifically, the slope of the cluster initial mass function appears to be correlated to the stellar mass of the galaxy, with the more massive galaxies having shallower slopes. Massive clusters ($M_\mathrm{cl}\geq6\times10^4 \Msun$) are rarely formed in galaxies with present-day stellar mass below $\Mstar \sim 10^9\Msun$.
\item Certain stellar feedback mechanisms in metal-rich clouds of gas tend to hinder the formation of star clusters. As such, massive clusters generally form at epochs when the metallicity of the galaxy is still sub-solar. The formation of very massive and dense GMCs, resulting mainly from major mergers of massive galaxies, can explain the presence of metal-rich clusters in MW like galaxies.
%This effect, combined with the general expectation that metallicity plays a role against the formation of merging BBHs, places certain constraints on the population of clusters that could be the birthplaces of observed BBHs.
\item In the most massive dwarf galaxies ($\Mstar \gtrsim 10^9\Msun$), a small number of very massive clusters, which are particularly efficient at forming merging BBHs, can make up for a similar contribution to the total population of merging BBHs as the less massive ones. In galaxies with masses similar to the MW or even more massive, one extremely massive and dense cluster may be sufficient to dominate the formation of merging BBHs over all the other massive clusters.
\item In agreement with previous studies, we find that the physical properties of BBHs formed in massive clusters can have very distinct features. The possibility of hierarchical mergers in such dense environments allow for the formation of BHs above the pair-instability mass gap, which are often part of BBHs with unequal mass components. These extreme events, which are hard to explain with isolated binary evolution, could be unique signatures of dynamical formation.
\item In addition to the general scaling with the galaxy stellar mass, there are clear correlations between the star formation history of each galaxy and the BBH merger rate from massive clusters. In particular, major mergers of galaxies appear to drive the formation of massive clusters, which in turn result in the formation of merging BBHs.
\end{enumerate}

This study represents the first attempt to estimate the populations of merging BBHs formed in massive clusters within cosmological simulations of galaxies. These results highlight the importance of modeling the formation of merging BBHs at a galactic level.
In a follow-up study, we will explore different formation channels of BBHs in these realistic environments and use constraints from the ever-growing list of BBH mergers reported by the LVK Collaboration to infer their respective contributions.

\begin{acknowledgements}
      Tristan Bruel and Astrid Lamberts are supported by the ANR COSMERGE project, grant ANR-20-CE31-001 of the French Agence Nationale de la Recherche.  Carl L.~Rodriguez acknowledges support from NSF Grant AST-2009916, NASA ATP Grant 80NSSC22K0722, a Alfred P.~Sloan Research Fellowship, and a David and Lucile Packard Foundation Fellowship.  This work was supported by the ‘Programme National des Hautes Energies’ (PNHE) of CNRS/INSU co-funded by CEA and CNES’ and the authors acknowledge HPC ressources from ‘Mesocentre SIGAMM’ hosted by Observatoire de la C\^ote d’Azur. This work made use of infrastructure services provided by the Science IT team of the University of Zurich (\url{www.s3it.uzh.ch}). Support for MYG was provided by NASA through the NASA Hubble Fellowship grant \#HST-HF2-51479 awarded  by  the  Space  Telescope  Science  Institute,  which  is  operated  by  the   Association  of  Universities  for  Research  in  Astronomy,  Inc.,  for  NASA,  under  contract NAS5-26555. 
\end{acknowledgements}


\bibliographystyle{aa} % style aa.bst
\bibliography{references}

\end{document}